\documentclass[12pt]{article}
\usepackage{amsmath,amsfonts,amssymb,latexsym}

\def\be{\begin{equation}}
\def\ee{\end{equation}}
\def\bq{\begin{eqnarray}}
\def\eq{\end{eqnarray}}
\def\beq{\begin{eqnarray*}}
\def\eeq{\end{eqnarray*}}

\begin{document}

\title{{\Huge Geodesics at Sudden Singularities}}
\author{{\Large John D. Barrow$^{1}$ and S. Cotsakis$^{2}$ } \\
%EndAName
$^{1}$DAMTP, Centre for Mathematical Sciences, \\
Cambridge University, Cambridge CB3 0WA, UK \\
$^{2}$Research group of Geometry, Dynamical Systems and Cosmology,\\
University of the Aegean, Karlovassi 83200, Samos, Greece.}
\maketitle

\begin{abstract}
\noindent We show that a general solution of the Einstein equations that
describes approach to an inhomogeneous and anisotropic sudden spacetime singularity does not experience
geodesic incompleteness. This generalises the result established for
isotropic and homogeneous universes. Further discussion of the weakness of
the singularity is also included.

\noindent PACS number: 98.80.-k
\end{abstract}

\section{\protect\bigskip Introduction}

There has been strong interest in the structure and ubiquity of finite-time
singularities in general-relativistic cosmological models since they were
first introduced by Barrow et al \cite{jb1}, as a counter-example to the
belief \cite{ellis} that closed Friedmann universes obeying the strong
energy condition must collapse to a future singularity. They were
characterised in detail as sudden singularities in\ refs. \cite{jb2, jb3,
jb4} and are `weak' singularities in the senses defined by Tipler \cite{tip}
and Krolak \cite{kr}. A sudden future singularity at $t_{s}$ is defined
informally in terms of the metric expansion scale factor, $a(t)$ with $%
t_{s}>0,$ by $0<a(t_{s})<\infty ,$ $0<\dot{a}(t_{s})<\infty ,\ddot{a}%
(t\rightarrow t_{s})\rightarrow -\infty $. These archetypal examples have
finite values of the metric scale factor, its first time derivative and the
density at a finite time but possess infinities in the second time
derivative of the scale factor and in the pressure. Higher-order examples
exist with infinities in the $(2+n)^{th}$ derivatives of the scale factor
and the $n^{th}$ derivative of the matter pressure \cite{jb3, jb4}. Other
varieties of finite-time singularity have been found in which a different
permutation of physical quantities take on finite and infinite values.%
\footnote{%
There is an interesting example in Newtonian mechanics of motion which
formally begins from rest with infinite acceleration. It is motion at
constant power. This means $v\dot{v}$ is constant, where $v=\dot{x}$ is the
velocity in the $x$ direction and so $v\propto t^{1/2}$ and $x\propto t^{3/2}
$ if initially $v(0)=x(0)=0$. Thus we see that the acceleration formally has
$\dot{v}\propto t^{-1/2}$ and diverges as $t\rightarrow 0$. This motion at
constant power is an excellent model of drag-car racing. The singularity in
the acceleration as $t\rightarrow 0$ is ameliorated in practice by the
inclusion of frictional effects on the initial motion \cite{drag}.}.

The general isotropic and homogeneous  approach to a sudden
finite-time singularity introduced in \cite{jb2} for the Friedmann universe has been
used \cite{bct} to construct a quasi-isotropic, inhomogeneous series
expansion around the finite-time singularity which contains nine
independently arbitrary spatial functions, as required of a part of the
general cosmological solution when the pressure and density are not related
by an equation of state. The stability properties of a wide range of
possible finite-time singularities were also studied in ref \cite{lip}.

It has also been shown by Fern\'{a}ndez-Jambrina and Lazkoz \cite{jambr,
jambr2, jambr3} that, in the context of the Friedmann universe, the sudden
singularity introduced in \cite{jb2} has the property that geodesics do not
feel the sudden singularity and pass through it. In this note we will
examine the evolution of geodesics in the general nine-function solution in
the vicinity of an inhomogeneous and anisotropic sudden singularity to see
if this result continues to hold. We will also formulate these earlier
results more precisely.

We will use Latin indices for spacetime components, Greek indices for space
components, and set $G=c=1$.

\section{Geometric setup}

Let $\Sigma _{0}$ be the 3-space defined by the equations $x^{i}=\phi
^{i}(\xi ),\xi =(\xi _{1},\xi _{2},\xi _{3})$, located at $t=0$. We suppose
that the sudden singularity is located at the time $t_{s}$ to the future,
and denote by $\Sigma _{s}$ the 3-space $t=t_{s}$. We may attach geodesic
normal (synchronous) coordinates at any point $B\in \Sigma _{s}$ as follows.
Let $u^{i}(\xi )$ be a $\mathcal{C}^{0}$ vector field over $\Sigma _{0}$,
and through any point on $\Sigma _{0}$ we draw causal geodesics tangent to $%
u^{i}(\xi )$ in both future and past directions parametrized by $t$. These
geodesics have $dx^{i}/dt=u^{i}$ (and $t=0$ on $\Sigma _{0}$). Then the
geodesic $x^{i}(t)$ that passes through $B$ cuts $\Sigma _{0}$ at the point $%
A$ with coordinates $(\xi _{1},\xi _{2},\xi _{3})$ where $t=0$ and $%
dx^{i}/dt=u^{i}$. The coordinates of $B$ are then $(t_{s},\xi )$, where $%
t_{s}$ is $t$ evaluated at $B$ and $\xi $ at $A$.

\section{$\mathcal{C}^1$ quasi-isotropic metric}

In \cite{bct} we found that near a sudden singularity the general form of
the metric in geodesic normal coordinates is
\begin{equation}
ds^{2}=dt^{2}-\gamma _{\alpha \beta }dx^{\alpha }dx^{\beta },\quad \gamma
_{\alpha \beta }=a_{_{\alpha \beta }}+b_{_{\alpha \beta }}t+c_{_{\alpha
\beta }}t^{n}+\cdots ,\quad n\in (1,2),  \label{met}
\end{equation}%
and the leading orders of the energy-momentum tensor components, defined by
\begin{equation*}
T_{j}^{i} =(\rho +p)u^{i}u_{j}-p\delta _{j}^{i}, \quad u_{a}u^{a} =1,
\end{equation*}%
are
\begin{equation}
u_{\alpha }=-\frac{3(b_{\alpha ;\beta }^{\beta }-b_{;\alpha })}{2n(n-1)c}%
\;t^{2-n}\sim t^{2-n},\quad u^{\alpha }=\gamma ^{\alpha \beta }u_{\beta
}\sim t^{2},
\end{equation}
\begin{equation*}
16\pi \rho =\left( P+\frac{b^{2}-b^{\mu \nu }b_{\mu \nu }}{4}\right) -\frac{n%
}{2}(b^{\mu \nu }c_{\mu \nu }-bc)\;t^{n-1}+\cdots ,
\end{equation*}
\begin{equation*}
16\pi p=-\frac{2n(n-1)c}{3}\;t^{n-2}-\frac{3b^{\mu \nu }b_{\mu \nu }+b^{2}+4P%
}{12}-\frac{n}{2}(b^{\mu \nu }c_{\mu \nu }+\frac{bc}{3})\;t^{n-1}+\cdots .
\end{equation*}
The Ricci scalar is
\begin{equation*}
R=R_{i}^{i}=-n(n-1)ct^{n-2}-\frac{b_{\mu \nu }b^{\mu \nu }+b^{2}+4P}{4}-%
\frac{n}{2}(b^{\mu \nu }c_{\mu \nu }+bc)\;t^{n-1}+\cdots ,
\end{equation*}
where $P$ is the trace of $P_{\alpha \beta }$, the spatial Ricci tensor
associated with $a_{\alpha \beta }.$

This solution is only $\mathcal{C}^{1}$, meaning the the metric, its first
derivatives as well as the Christoffel symbols will be continuous through
the 3-slice $\Sigma _{s}$ containing the sudden singularity at $B$, but we
expect discontinuities in the second and higher derivatives of the metric,
and at least in the first derivatives of the Christoffel symbols.

\section{Geodesic behaviour at $t_s$}

The Christoffel symbols are $\mathcal{C}^{0}$, and so the geodesic
equations,
\begin{equation}
\ddot{x}^{i}+\Gamma _{jk}^{i}u^{j}u^{k}=0,  \label{g}
\end{equation}%
will have solutions, $x^{i}(t)$, with continuous derivatives up to and
including ${d^{2}x^{i}}/{dt^{2}}$. Therefore, we can Taylor estimate these
solutions as follows. For any $\delta >0$ and $t\in (t_{s}-\delta
,t_{s}+\delta )$, we have
\begin{equation}
x^{i}(t)=x^{i}(t_{s})+(t-t_{s})u^{i}(t_{s})-\frac{1}{2}(t-t_{s})^{2}(\Gamma
_{\alpha \beta }^{i}u^{\alpha }u^{\beta })(t_{\ast }),  \label{exp}
\end{equation}%
with $t_{\ast }$ between $t$ and $t_{s}$. The last term is given in the
Lagrange form for the remainder. Since the error term is quadratic in $%
t-t_{s}$, it vanishes asymptotically for both past and future sudden
singularities. This means that the geodesic equations (\ref{g}) have
complete $\mathcal{C}^{2}$ solutions through the sudden singularity at $B$
to the future and the past given by this form. In higher-order lagrangian
theories of gravity it is possible for sudden singularities to arise because
there are infinities in the third, or higher, time derivatives of the metric
scale factor. In these cases the effect of the singularity on the geodesics
is weaker still and avoids a violation of the dominant energy condition \cite%
{lake, jb4}.

A spacetime is Tipler(T)-strong \cite{tip} iff, as the affine parameter $\tau
\rightarrow t_{s}$, the integral
\begin{equation}
T(u)\equiv \int_{0}^{\tau }d\tau ^{\prime }\int_{0}^{\tau ^{^{\prime
}}}R_{ij}u^{i}u^{j}d\tau ^{\prime \prime }\rightarrow \infty .
\end{equation}%
The spacetime is Krolak(K)-strong \cite{kr} iff, as $\tau \rightarrow t_{s}$,
the integral
\begin{equation}
K(u)\equiv \int_{0}^{\tau }R_{ij}u^{i}u^{j}d\tau ^{\prime }\rightarrow
\infty .
\end{equation}%
If these conditions do not hold the spacetime is T-weak or K-weak,
respectively. It is possible for a singularity to be K-strong but T-weak,
for example the so-called \cite{type} Type III singularities with $\rho
\rightarrow \infty ,\left\vert p\right\vert \rightarrow \infty $ as $%
a\rightarrow a_{s}$ have this property. In our case, the various components
of the Ricci curvature have leading orders of the following forms: $%
R_{00}\sim t^{n-2},R_{0\alpha }\sim t^{0},R_{\alpha \gamma }\sim t^{2(n-1)},$
while $u^{0}\sim t^{0},u^{\alpha }\sim t^{2}$. Therefore
\begin{equation}
R_{ij}u^{i}u^{j}\sim t^{n-2}+2t^{2}+t^{2n+2}.
\end{equation}%
But since at the sudden singularity, $1<n<2$, we find that
\begin{equation}
R_{ij}u^{i}u^{j}\sim t^{n-2},\quad \text{as}\quad t\rightarrow t_{s},
\end{equation}%
and so after one integration we have,
\begin{equation}
K(u)\sim \tau ^{n-1}\rightarrow t_{s}^{n-1},\quad \text{as}\quad \tau
\rightarrow t_{s},
\end{equation}%
and after a second integration,
\begin{equation}
T(u)\sim \tau ^{n}\rightarrow t_{s}^{n},\quad \text{as}\quad \tau
\rightarrow t_{s},
\end{equation}%
and so the generic sudden singularity (\ref{met}) is T-weak and K-weak%
\footnote{%
If $0<n<1,$ and the metric contains a power of $(t-t_{s}\ )^{n}$, then it
will only have a well-defined meaning as a real function when $t-t_{s}>0$.
So, at any point $t_{s}$ (e.g., when $t_{s}=0$), it will be defined
asymptotically only in the past direction and not to the future. Our
argument~also requires continuity of the Christoffel symbols, and so it will
not be valid when the metric contains a power of $(t-t_{s}\ )^{n}$ with $%
0<n<1$, even if we restrict only to the past direction. For an isotropic
solution with a sudden singularity at $t=0$, see \cite{dab}}. This weakness
also suggests that we do not expect these singularity structures to be
modified by quantum particle production effects. Some studies of the quantum
cosmology of sudden singularities which confirm this have been made in refs
\cite{quan} but quantum modifications can occur for particular
regularisation procedures \cite{haro}. There are also interesting classical
questions about the passage through a sudden singularity in certain examples
where the background matter variables, $\rho $ and $p$, do not continue to
be well defined. These problems can be avoided by a distributional
redefinition of the cosmological quantities involved \cite{dist}. It is also
interesting to note that extended objects like fundamental string loops can
pass through weak singularities without their invariant sizes becoming
infinite \cite{bal}.

\section{Conclusion}

This result generalizes the studies of Fern\'{a}ndez-Jambrina and Lazkoz
\cite{jambr, jambr2, jambr3} by showing that there is no geodesic
incompleteness at a general inhomogeneous and anisotropic sudden
singularity. The inclusion of anisotropy and inhomogeneity does not
introduce geodesic incompleteness. We would expect that these results will
also hold for sudden singularities in Loop Quantum Gravity cosmologies of
the sort studied in ref. \cite{singh} and in higher-order lagrangian gravity
theories \cite{jb3}.

\end{document}